\documentclass[]{JHEP3}
\usepackage{epsfig,multicol,bbm,diagbox,makecell}

\newcommand\fverb{\setbox\fverbbox=\hbox\bgroup\verb}
\newcommand\fverbdo{\egroup\medskip\noindent%
                        \fbox{\unhbox\fverbbox}\ }
\newcommand\fverbit{\egroup\item[\fbox{\unhbox\fverbbox}]}
\newbox\fverbbox

\title{Prediction for CP Violation  via Electric Dipole Moment of $\tau$ Lepton
 in ${\gamma\gamma \to \tau^{+}\tau^{-}}$ Process at CLIC } 

\author{S. Ata\u{g}\\
Department of Physics, Faculty of Sciences,
Ankara University, 06100 Tandogan, Ankara, Turkey\\
E-mail:\email{atag@science.ankara.edu.tr}}

\author{E. G\"{u}rkanl\i \\
\thanks{Also at Department of Physics, Faculty of Sciences,
Ankara University, 06100 Tandogan, Ankara, Turkey}
Department of Physics, Sinop University,
57000, Sinop, Turkey\\
E-mail:\email{gurkanli@science.ankara.edu.tr}}

\abstract{Pair production of tau leptons in 
two photon collision  
${\gamma\gamma \to \tau^{+}\tau}^{-}$ 
is studied at CLIC to test CP violating
QED couplings of tau leptons.  
CP violating effects are 
investigated using tau pair spin correlations 
which are observed through the hadronic decay 
of each $\tau$ into $\pi\nu$. 
Competitive bounds with previous works 
on the electric dipole moment from  CP odd terms
have been obtained.}


\begin{document}

\section{Introduction}

Although several recent results,  including discovery of Higgs boson 
have been obtained from Large Hadron Collider (LHC), there has still
been physics unexplored concerning CP violation. Information from Standard 
Model (SM) is not enough to have deeper understanding of the origin of CP violation.
In leptonic interactions there is no CP violating couplings but
it is possible that multi-loop  contributions 
from quark sector \cite{ckm,hoog} indirectly induce CP violation which is too small to detect.
Extensions of the Standard Model which cover neutrino mixing with different 
neutrino masses \cite{weinberg}, more  Higgs multiplets \cite{barr} 
could introduce CP violating couplings  
into the lepton sector. One of the motivations to go beyond the Standard Model 
comes from the baryon asymmetry in the universe. Supersymmetry \cite{ellis}, 
Left-Right  Symmetric models \cite{pati}, 
leptoquark models \cite{ma}  have been proposed for additional 
 sources of CP   violation. 
There is no detectable electric dipole form factors of leptons in the Standard Model.
If one considers coupling of leptons beyond the Standard Model, 
electric dipole  form factors  may cause detectable size of CP violation.
Tau lepton is expected to have larger effects because of its large mass.
 In this work, we are interested in   CP violating 
effects using the following parametrization for the 
effective electromagnetic   coupling of tau lepton at the vertex $\gamma\tau\tau$
 as an addition to electric charge  $\gamma^{\mu}$ coupling \cite{grimus}:
  
\begin{eqnarray}
 \Gamma^{\mu}=&& F_{1}(q^2)\gamma^{\mu}+
+F_{2}(q^{2})\frac{i}{2m}\sigma^{\mu\nu}q_{\nu}
+F_{3}(q^{2})\frac{1}{2m}\sigma^{\mu\nu}q_{\nu}\gamma^{5} \\
\sigma^{\mu\nu}=&&\frac{i}{2}(\gamma^{\mu}\gamma^{\nu}-
\gamma^{\nu}\gamma^{\mu})
\end{eqnarray}
where q and m  are the momentum transfer to the photon and the mass of tau lepton.
$F_{1}(q^{2})$, $F_{2}(q^{2})$ and  $F_{3}(q^{2})$ are electric charge,
 anomalous magnetic dipole and   electric dipole form factors.  
In the limiting case of $q^{2}\to 0$,  the form factors are called moments 
which describe   the static properties of the fermions.
\begin{eqnarray}
F_{1}(0)=1,\,\,\, a_{\tau}=F_{2}(0), \,\,\,
d_{\tau}=\frac{e}{2m}F_{3}(0)
\end{eqnarray}
In the case of electron and muon, anomalous electromagnetic moments have been 
measured or constrained with high accuracy at low energy spin precession 
experiments. Because of the higher mass and short life time, experiments 
with tau lepton need colliders. This causes additional uncertainty  in 
identification of tau lepton during production and decay processes  
when compared to low mass leptons. Accurate SM calculation  
with three loops gives  tau lepton electric dipole moment (EDM)
 of the order of $10^{-35}$ e cm \cite{ckm,hoog}.
Since this result is far from the  present experimental capability, any observation
at colliders about tau EDM  shows the indication of 
new physics beyond the Standard Model.   
CP violating electric dipole form factor leads to contribution to the 
cross sections proportional to $F_{3}^2$ or higher order  even power 
in many cases. CP even terms in cross sections can not be considered as a test 
of CP violation. A true CP violating  contribution should appear 
linear or  odd power  in 
$F_{3}$ in the cross section which results from  the interference 
with the Standard Model amplitude. 

Upper limits on the electric dipole moment of the $\tau$ lepton 
with CP even terms in the cross sections have been
obtained so far from the experiments at LEP  
\cite{l3,opal,delphi}
\begin{eqnarray}
|d_{\tau}| &&< 3.1\times 10^{-16}\,\, \mbox{e cm}\,\, \mbox{(L3)} \\
|d_{\tau}| &&< 3.7\times 10^{-16}\,\, \mbox{e cm}\,\, \mbox{(OPAL)} \\
|d_{\tau}| &&< 3.7\times 10^{-16}\,\, \mbox{e cm}\,\, \mbox{(DELPHI)}.
\end{eqnarray}
BELLE Collaboration used spin correlation observables to obtain 
limits on the electric dipole moment with CP odd terms 
in the cross section through  the process 
$e^{+}e^{+} \to \gamma \to \tau^{+}\tau^{-}$
\cite{belle}  
\begin{eqnarray}
-0.22<\mbox{Re}(d_{\tau}) < 0.45 \,\,( 10^{-16}\,\, \mbox{e cm}) \\
-0.25<\mbox{Im}(d_{\tau}) < 0.08 \,\,( 10^{-16}\,\, \mbox{e cm}). 
\end{eqnarray}
In this process, the intermediate photon has virtualtiy of
$Q^2=100$ $GeV^2$. Then, their bounds on real and imaginary parts
of $d_{\tau}$ were given separately. 
There are several articles providing limits on electric dipole moment 
from previous LEP results  or   
by using some indirect methods. Here we are going to mention 
only about the articles giving results based on  terms linear in $d_{\tau}$
or CP odd terms  in the cross sections \cite{vidal}.  

In this paper, we are going to discuss the potential of CLIC to constrain 
EDM of tau lepton with CP odd terms in the cross section via tau pair 
production process in two photon collision ${\gamma\gamma \to \tau^{+}\tau}^{-}$.
Direct production of tau  pair  $e^{+}e^{-} \to \tau^{+}\tau^{-}$ 
has   small cross section due to s-channel 
Feynman amplitude with CLIC energies 500-3000 GeV in comparison to 
BELLE with center of mass energy around 10 GeV Fig.\ref{fig1}-a. 
Another disadvantage of direct production is connected to the photon 
virtuality $Q^2=10^{5}-10^{7}$ $GeV^2$ which is not convenient for 
the definitions of electromagnetic moments with $Q^2 \to 0$ .
Pair production of charged leptons in
two photon collision provide  a unique opportunity to test
Quantum Electrodynamics (QED). This process is followed by
$e^{+}e^{-} \to e^{+}e^{-} \tau^{+}\tau^{-}$ via t-channel subprocess 
$\gamma\gamma \to \tau^{+}\tau^{-} $ based on equivalent photon 
approximation (later sections)  Fig.\ref{fig1}-b. Additional benefit from 
two photon collision  is the proper photon
virtuality $Q^2 \to 0$  at two vertices  containing  anomalous couplings.
In this case, contrary to s-channel, t-channel contribution to the cross section 
increases with increasing energy due to sum over photon spectrum . 
As will be shown later, direct production cross sections are
smaller than the case of two photon induced production of tau pair by a factor of 
$10^{-3}-10^{-4}$ for CLIC energies.

The spin averaged cross section brings even powers of $F_{3}$ in the result that 
means CP is even. CP odd terms appear if spin dependent cross section is taken 
into account. Then we will need an observable which selects only CP odd terms      
from the cross section.  
In the next section, we give  some details of the spin correlated 
cross section in the sub process $\gamma\gamma \to \tau^{+}\tau^{-}$  
with the anomalous couplings  of the $\tau$ lepton at both vertices.
In section 3 we define  azimuthal asymmetry as the spin correlated observable
in terms of the final decay products of tau leptons.  Section 4 deals with     
the equivalent photon approximation and 
the bounds on the electric dipole moment obtained for CLIC parameters. 
In the last section, we discuss the points  concerning the background 
and tau lepton identification. 
\FIGURE{\epsfig{file=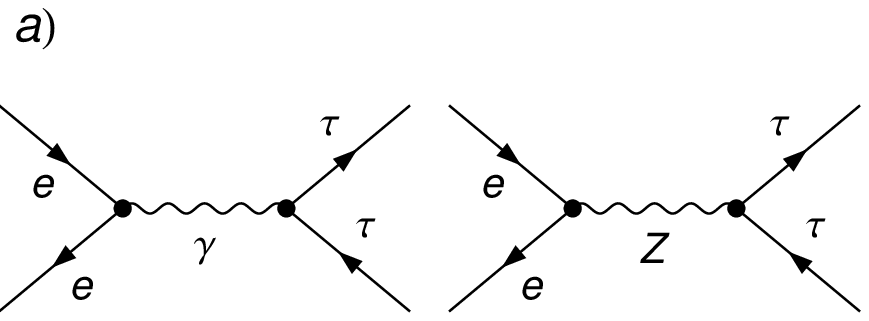} 
\epsfig{file=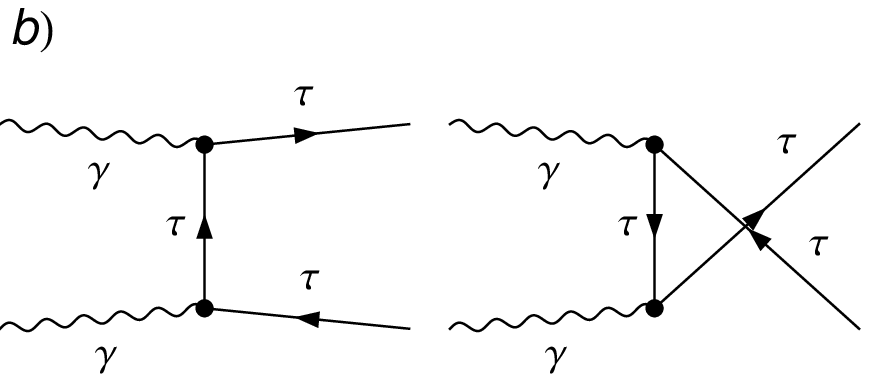 }
\caption{a)Direct production of tau pair at CLIC. b) Two photon induced 
tau pair production with equivalent photon approximation. }
\label{fig1}}

\section{Spin Correlated Cross Section For Tau Pair production}

For the subprocess $\gamma\gamma \to \tau^{+}\tau^{-}$,
Feynman diagrams of t and u channels, as seen in Fig.\ref{fig1}-b,  
are responsible where anomalous couplings are included by both vertices. 
Center of mass system
of tau pairs is taken into account for calculation. Since spin dependent cross 
section is required, convenient coordinate system with z-axis along the produced 
$\tau^{-}$ lepton is chosen. In this frame,     spin vectors of  tau leptons   
in four dimensions can be given as
\begin{eqnarray}
s_{-}^{\mu}=&&(\frac{p_{1z}}{m}s_{1z}, s_{1x},s_{1y},\frac{E}{m}s_{1z} ) \\
s_{+}^{\mu}=&&(\frac{-p_{1z}}{m}s_{2z}, s_{2x},s_{2y},\frac{E}{m}s_{2z} )
\end{eqnarray}
where $\vec{s}_{1}$ and $\vec{s}_{2}$ are spin vectors of tau leptons
 in the rest frame of each tau. $p_{1z}$, and  E are momentum and energy  of  
$\tau^{-}$, respectively.    
The squared amplitude are written explicitly in terms of  reduced amplitudes and
tau spins  with  cartesian components \cite{tsai}
\begin{eqnarray}
|M_{1}|^{2}&&=\frac{16\pi^{2}\alpha^{2}}{(\hat{t}-m^{2})^{2}}
[C_{0}+C_{xx}s_{1x}s_{2x}+C_{yy}s_{1y}s_{2y}+C_{zz}s_{1z}s_{2z}\nonumber\\
&&+C_{xy}^{-} (s_{1x}s_{2y}-s_{1y}s_{2x})\nonumber\\
&&+C_{yz}^{-} (s_{1y}s_{2z}-s_{1z}s_{2y})
+C_{xz}^{+} (s_{1x}s_{2z}+s_{1z}s_{2x})] \\
|M_{2}|^{2}&&=\frac{16\pi^{2}\alpha^{2}}{(\hat{u}-m^{2})^{2}}
[D_{0}+D_{xx}s_{1x}s_{2x}+D_{yy}s_{1y}s_{2y}+D_{zz}s_{1z}s_{2z}
\nonumber\\&&+D_{xy}^{-} (s_{1x}s_{2y}-s_{1y}s_{2x})\nonumber\\
&&+D_{yz}^{-} (s_{1y}s_{2z}-s_{1z}s_{2y})
+D_{xz}^{+} (s_{1x}s_{2z}+s_{1z}s_{2x})] \\
|M_{12}|+|M_{21}|&&=\frac{16\pi^{2}\alpha^{2}}{(\hat{t}-m^{2})
(\hat{u}-m^{2})}[G_{0}+G_{xx}s_{1x}s_{2x}+G_{yy}s_{1y}s_{2y}+G_{zz}s_{1z}s_{2z}
\nonumber\\&&+G_{xy}^{-} (s_{1x}s_{2y}-s_{1y}s_{2x})
\nonumber\\&&+G_{yz}^{-} (s_{1y}s_{2z}-s_{1z}s_{2y})
+G_{xz}^{+} (s_{1x}s_{2z}+s_{1z}s_{2x})]
\end{eqnarray}
where  definitions of the quantites  $C_{ij},D_{ij},G_{ij}$, 
$i,j=0,x,y,z$  are given in the Appendix.
In above expressions  $k_{1}$, $k_{2}$, $p_{1}$ and $p_{2}$  are the 
momenta of the incoming photons and final $\tau$ leptons. 
Mandelstam variables are defined as $\hat{s}=(k_{1}+k_{2})^{2}$,
$\hat{t}=(k_{1}-p_{1})^{2}$ and $\hat{u}=(k_{1}-p_{2})^{2}$. 
In the terms  $C_{ij},D_{ij},G_{ij}$ there are $F_{3}F_{1}^{3}$ ,
$F_{3}F_{2}F_{1}^{2}$, $F_{3}^{3}F_{1}$ and terms with  even powers of $F_{3}$.
Azimuthal asymmetry (next section) will select odd powers of $F_{3}$ terms in the 
numerator. However, the dominant term will be  $F_{3}F_{1}^{3}$  where $F_{1}=1$ is 
pointlike coupling, then we have retained $F_{3}$ and $F_{3}^{2}$ parts
 and higher orders  $F_{3}^{3}$, $F_{3}^{4}$ have been neglected due to their 
smallness. There is also a term $F_{2}F_{1}^{3}$ that does not  contribute 
to azimuthal asymmetry.    
We should notice that $\vec{s}_{1}$ and
$\vec{s}_{2}$ appear bilinearly which means two spins are correlated.   
Some of  terms with anomalous coupling have spin correlations too. The terms
proportional to $F_{3}$ are CP odd and spin correlated. It is interesting 
that when  spin correlation is removed  CP odd terms vanish. Then we need convenient
observables to determine size of CP violation within this process.    
When tau pair decays into charged particles and neutrinos, measuring spin correlation 
transforms to  angular distributions of decay products.  
First we examine tau decay rate independently from pair production
  in the frame where tau lepton has momentum $p_{1}$. 
For the decay mode to a charged hadron and a neutrino $\tau \to h+\nu$ 

\begin{eqnarray}
\frac{1}{\Gamma}\frac{d\Gamma}{d\Omega_{h}}=\frac{1}{4\pi}\frac{m^2}
{(E-\vec{p}_{1}\cdot\hat{p}_{h})^2}[1-\vec{V}_{h}\cdot\vec{s}_{1}]
\end{eqnarray}
with 

\begin{eqnarray}
\vec{V}_{h}=\frac{1}{(E-\vec{p}_{1}\cdot\hat{p}_{h})}[(1-\frac{\vec{p}_{1}
\cdot\hat{p}_{h}}{E+m})\vec{p}_{1}-m\hat{p}_{h}]
\end{eqnarray}
 where $\hat{p}_{h}$ shows unit momentum vector of hadron such as charged pion. From here on 
 pion masses will be neglected. If the rest frame of tau is needed, 
$\vec{p}_{1}$ is taken to be zero and one gets familiar 
result 
\begin{eqnarray}
\frac{1}{\Gamma}\frac{d\Gamma}{d\Omega_{h}}=\frac{1}{4\pi}[1+\alpha\hat{p}_{h}\cdot\vec{s}_{1}]
,  \,\,\,\,\,\, \alpha=\pm 1 \,\,\, \mbox{for charged pions} .
\end{eqnarray}

In the subprocess $\gamma\gamma\to \tau^{-}\tau^{+}\to (h^{-}\nu)( h^{+}\bar{\nu})$
 tau decays occur through tau propagators. Around resonance, narrow width approximation 
is convenient to discuss  correlation between decay products caused by tau spins.
It is possible to combine tau pair production  and decay parts to give following form
\cite{tsai}

\begin{eqnarray}
\frac{d\hat{\sigma}}{d\cos\theta}=\frac{BR(\tau^{-} \to h^{-}\bar{\nu})
BR(\tau^{+} \to h^{+}\nu)}{16\pi^2}
\int d\Omega_{h^{-}} d\Omega_{h^{+}}
[A + \sum_{i,j}B_{ij}\, V_{h^{-}}^{i}V_{h^{+}}^{j} ] .
\end{eqnarray}
The indices $i,j$ indicate cartesian components x,y,z. The angle $\theta$ is between 
tau lepton and incoming photon in the center of mass system of tau pair.
The contents of $A$, one of $B_{ij}$ and relations between them are given below

\begin{eqnarray}
A=&&\frac{\pi\alpha^2}{2\hat{s}}[\frac{C_{0}}{(\hat{t}-m^{2})^{2}}
+\frac{D_{0}}{(\hat{u}-m^{2})^{2}}-\frac{G_{0}}{(\hat{t}-m^{2})(\hat{u}-m^{2})}]\\
B_{xy}=&&\frac{\pi\alpha^2}{2\hat{s}}[\frac{C_{xy}^{-}}{(\hat{t}-m^{2})^{2}}
+\frac{D_{xy}^{-}}{(\hat{u}-m^{2})^{2}}-\frac{G_{xy}^{-}}{(\hat{t}-m^{2})(\hat{u}-m^{2})}]\\
B_{yx}=&&-B_{xy}\\
B_{zy}=&&-B_{yz}\\
B_{zx}=&&B_{xz} .
\end{eqnarray}

\section{Azimuthal Asymmetry }

In order to single out CP odd terms proportional to $F_{3}$ 
azimuthal asymmetry   definition can be made using the angles 
of decay products. Because tau momenta is in the z-direction,
azimuthal angles of decay products 
do not change  even if  the boost along tau momenta is applied. Then, integrations 
over azimuthal angles can be performed in a convenient reference system. 
 Remaining  integrations over polar angles 
 will be done using full angular regions.  
Therefore, we replace $V_{h^{-}}^{i}V_{h^{+}}^{j}$ 
by $\hat{p}_{h^{-}}^{i}\hat{p}_{h^{+}}^{j}$ for simplicity.
The terms proportinal to $F_{3}$ in the cross section appear as 
$B_{xy}(\hat{p}_{h^{-}}^{x}\hat{p}_{h^{+}}^{y}-\hat{p}_{h^{-}}^{y}\hat{p}_{h^{+}}^{x})$
with the following  momentum directions
\begin{eqnarray}
\hat{p}_{h^{-}}=&&(\sin\theta_{-} \cos\phi_{-}, \sin\theta_{-} \sin\phi_{-}, 
\cos\theta_{-} )\\
\hat{p}_{h^{+}}=&&(\sin\theta_{+} \cos\phi_{+}, \sin\theta_{+} \sin\phi_{+}, 
\cos\theta_{+} ).
\end{eqnarray}
The integration over azimuthal angles $\phi_{-}$ and  $\phi_{+}$ can be organized 
according to signature of $\delta=\sin(\phi_{-}-\phi_{+})$ based on the expression 
in the integrand below
\begin{eqnarray}
(\hat{p}_{h^{-}}^{x}\hat{p}_{h^{+}}^{y}-\hat{p}_{h^{-}}^{y}\hat{p}_{h^{+}}^{x})=
\sin\theta_{-}\sin\theta_{+}\sin(\phi_{-}-\phi_{+}).
\end{eqnarray}
It is required to  use  azimuthal asymmetry which compares number of  decay products 
scattered normal to the left and to the right regions of the production 
plane of tau pair as defined by signature of $\delta$  

\begin{eqnarray}
A_{az}=\frac{\int_{\delta>0}{d\sigma}-\int_{\delta<0}{d\sigma}}
{\int_{\delta>0}{d\sigma}+\int_{\delta<0}{d\sigma}} .
\end{eqnarray}
This asymmetry keeps only CP violating terms with $F_{3}$ in the numerator. 
After performing integrals over angular parts of decay products 
in the denominator, we are left with expression  without spin contributions
such as $C_{0},D_{0}$ and $G_{0}$ which include an additional terms 
 with $F_{3}^2$ and higher order even powers.
When compared with standard model part, we ignore   $F_{3}^2$ and  
higher even powers  in the denominator. 
While azimuthal asymmetry is taken into account, 
all angular integrations belong to decay products can be easily done in analytic way.     
Integrations over incoming photon spectrum and $\cos\theta$  belongs to tau pair production 
will be taken numerically.   

\section{Limits on CP Violation at CLIC }

CLIC  has  a potential to produce almost real gamma beam 
from electron and positron beams  with beam energy $E_{b}=250-1500$ GeV.   
Therefore, two photon induced  events at CLIC have large cross sections with 
final tau leptons. For additional integrations over each photon spectrum   
 we consider   equivalent photon approximation  based on the work \cite{budnev}

\begin{eqnarray}
\frac{dN}{dE_{\gamma}}=f(x)=&&\frac{\alpha}{\pi E_{b}}[(\frac{1-x+x^{2}/2}{x})
\log\frac{Q_{max}^{2}}{Q_{min}^{2}}-\frac{m_{e}^2 x}{Q_{min}^{2}}(1-
\frac{Q_{min}^{2}}{Q_{max}^{2}}) \\ \nonumber
-&&\frac{1}{x}(1-x/2)^2 \log\frac{x^2 E^{2}_{b}+Q_{max}^2}
{x^2 E^{2}_{b}+Q_{min}^{2}} ] 
\end{eqnarray} 
where $m_{e}$ is the electron mass,  $x=E_{\gamma}/E_{b}$  and $Q_{max}^2$ is the 
maximum virtuality. Kinematic definition of $Q_{min}^{2}$ can be given by 

\begin{eqnarray}
Q_{min}^{2}=\frac{m_{e}^2 x^2}{1-x} .
\end{eqnarray}
Using the above function for both beams 
  $f_{1}(x_{1})$ and $f_{2}(x_{2})$ with incoming photon energies 
$E_{1}$ and $E_{2}$, the cross section becomes
\begin{eqnarray}
d\sigma=&& \int dE_{1} dE_{2} f_{1}(x_{1})f_{2}(x_{2}) d\hat{\sigma}(\hat{s})\\
=&& \int \frac{dL^{\gamma\gamma}}{dW} d\hat{\sigma}(W) dW \\
\hat{s}=&& W^2 .
\end{eqnarray}  
Here W is the invariant mass of incoming photons and $dL^{\gamma\gamma}/dW $ is 
defined as  the effective photon luminosity
\begin{eqnarray}
\frac{dL^{\gamma\gamma}}{dW}=\int_{y_{min}}^{y_{max}} dy \frac{W}{2y}f_{1}
(\frac{W^2}{4y})f_{2}(y) 
\end{eqnarray} 

\begin{eqnarray}
y_{min}=\max(\frac{W^2}{4E_{2}^{max}},E_{1}^{min}), \,\,\,  y_{max}=E_{1}^{max} .
\end{eqnarray}
If scattered electrons of the beams are detected, maximum and minimum values of 
incoming photon energies  can be determined experimentally. 
 Otherwise  final energy or momentum cuts   of produced tau pair will be used 
to specify minimum photon energy. In order to get idea about the features of effective 
photon luminosity a graph of Eq.(4.6) has been drawn in Fig.\ref{fig2} 
for CLIC energies of $\sqrt{s}$=500 GeV, 3000 GeV  \cite{clic,clic2,clic3} 
and for $Q_{max}^2=2$ $\mbox{GeV}^2$.
 As seen from this figure, lower values of incoming
photon energies are much more effective on   the cross section. In fact, $\% 50$ of cross section
of the process can be reached by the energy $ W_{max}=0.03\sqrt{s}$. 
Thus, effective photon energy W for
the cross section   becomes less than 15-90 GeV depending on electron beam energy.

For small values of $Q_{max}^2$  and small collider energies, 
the last term in Eq.(4.1) is less effective and 
may be negligible  but 
small x values, large collider energies and  large $Q_{max}^2$   
the last term becomes effective and reduces the cross sections considerably. 
 There are several papers which use the code CompHep/CalcHep \cite{pukhov}
or similar packages to compute cross sections. All these packages use 
Weizsaecker-Williams  photon spectrum f(x) without the last term in Eq.(4.1).    
In order to see $Q_{max}^2$ dependence with and without third term 
we give   Standard Model $ee \to ee \tau\tau$ cross sections at CLIC
 in Table \ref{tab1}  for  different $Q_{max}^2$  values . It is understood that 
full photon spectrum leads smaller cross sections and 
smoother behaviour against $Q_{max}^2$. After imposing a transverse momentum cut 
on the final tau leptons the difference between the full and approximate 
photon spectrum gets smaller.
In order to make a comparison, Standard Model  cross sections of  
direct production  of tau lepton pair $ee\to \tau\tau$ are given below:
\begin{eqnarray}
&&\sigma=0.45 \,\mbox{pb}  \,\,\,  \mbox{for} \,\,\, \sqrt{s}=500 \, \mbox{GeV} \\
&&\sigma=0.01 \, \mbox{pb}  \,\,\,  \mbox{for} \,\,\, \sqrt{s}=3000 \, \mbox{GeV}
\end{eqnarray}
where direct production cross sections are smaller by a factor of
$10^{-3}-10^{-4}$ than the values shown in Table \ref{tab1}.   

\FIGURE{\epsfig{file=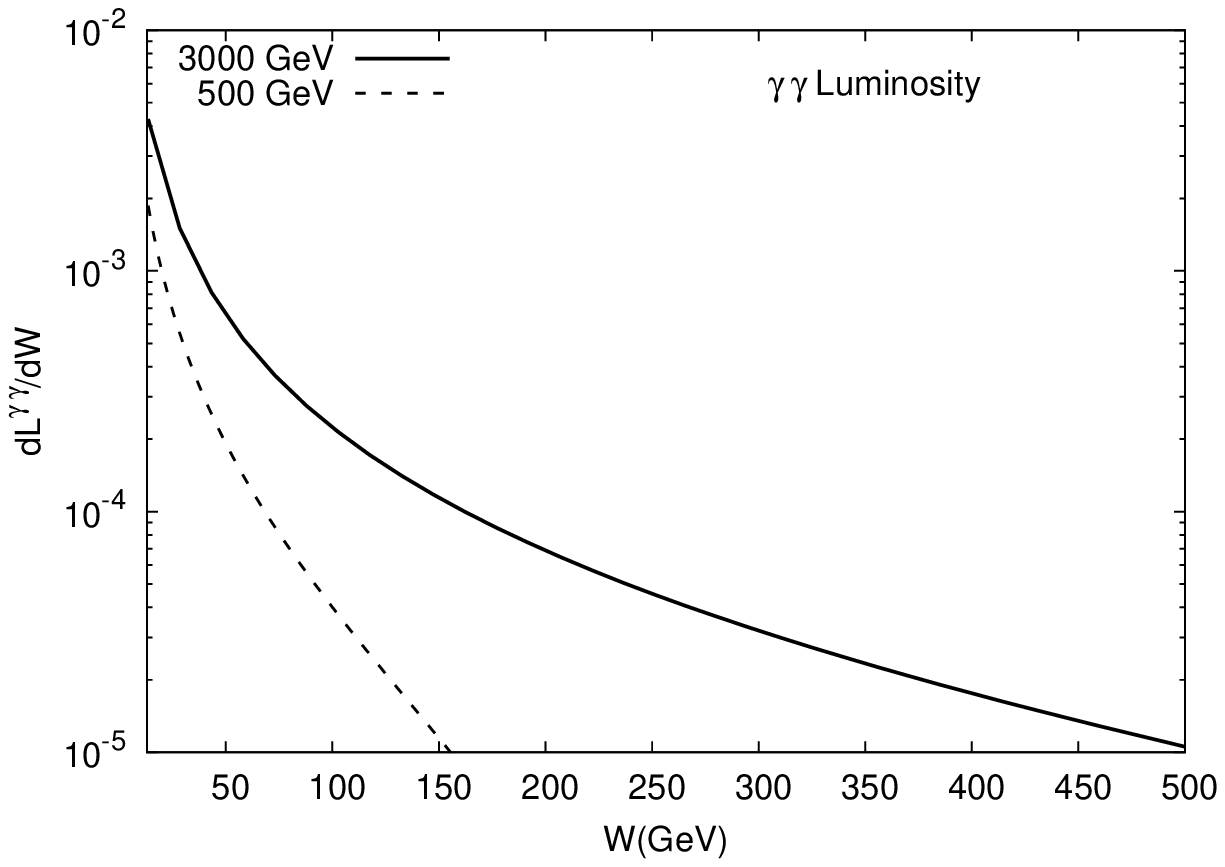}
\caption{Effective $\gamma\gamma$ luminosity as a function
of the invariant mass of the two photon system for  electron beam energies 
$\sqrt{s}=$ 500, 3000 GeV and $Q_{max}^2=2$ GeV$^{2}$}
\label{fig2}}

\TABLE{
\begin{tabular}{|c|c|c|c|}\hline
$Q_{max}^2$ (GeV$^2$) & $\sqrt{s}$ (GeV)& $\sigma_{SM}$(pb)
& $\sigma_{SM}^{PT}$(pb) \\
 \hline \hline
2   & 500 &   332 (360)  & 56(58)    \\
2   & 1500 &  523 (578)  & 99(103)    \\
2   & 3000 &  667 (744)  & 131(139)    \\
\hline
16   & 500 &   371 (428)  & 65(70)    \\
16   & 1500 &  581 (680)  & 112(123)    \\
16   & 3000 &  737 (869)  & 148(163)    \\
\hline
64   & 500 &   392 (478)  & 70(78)    \\
64   & 1500 &  613 (753)  & 120(136)    \\
64   & 3000 &  776 (958)  & 158(181)    \\
\hline
\end{tabular}
\caption{Standard Model cross sections of 
$ee\to ee \tau^{+}\tau^{-}$  for CLIC energies and maximum 
virtuality $Q_{max}^2$ values. The numbers in parenthesis
show cross sections obtained from the  photon spectrum 
neglecting the last term in Eq.(4.1).
In the last column   $ W_{min}=2\sqrt{m_{\tau}^2+P_{T}^2}$
have been considered  with a cut
$P_{T}=5$ GeV of each tau lepton. 
\label{tab1}}}

After having information about equivalent photon approximation, 
azimuthal asymmetry as a function of  $\cos{\theta}$ between tau leptons 
and incoming photons in the center of mass system of tau leptons 
has been obtained from Eq.(3.4) for tau decay products of $\pi^{\mp}$ $\nu$. 
During computations energy and transverse momentum cuts 0.2 GeV on each 
final state charged pion were applied.
Its properties are given by Fig.\ref{fig3} for 
$\sqrt{s}=500$ GeV and $\sqrt{s}=3000$ GeV. According to Fig.\ref{fig3},
highest asymmetry occurs when  tau leptons are produced transverse to 
incoming photons direction. If  integration over $\cos{\theta}$ 
is completed  we get almost the same asymmetry $A_{az}/F_{3}=0.70$ for the 
energies of $\sqrt{s}=500, 1500, 3000$ GeV

\FIGURE{\epsfig{file=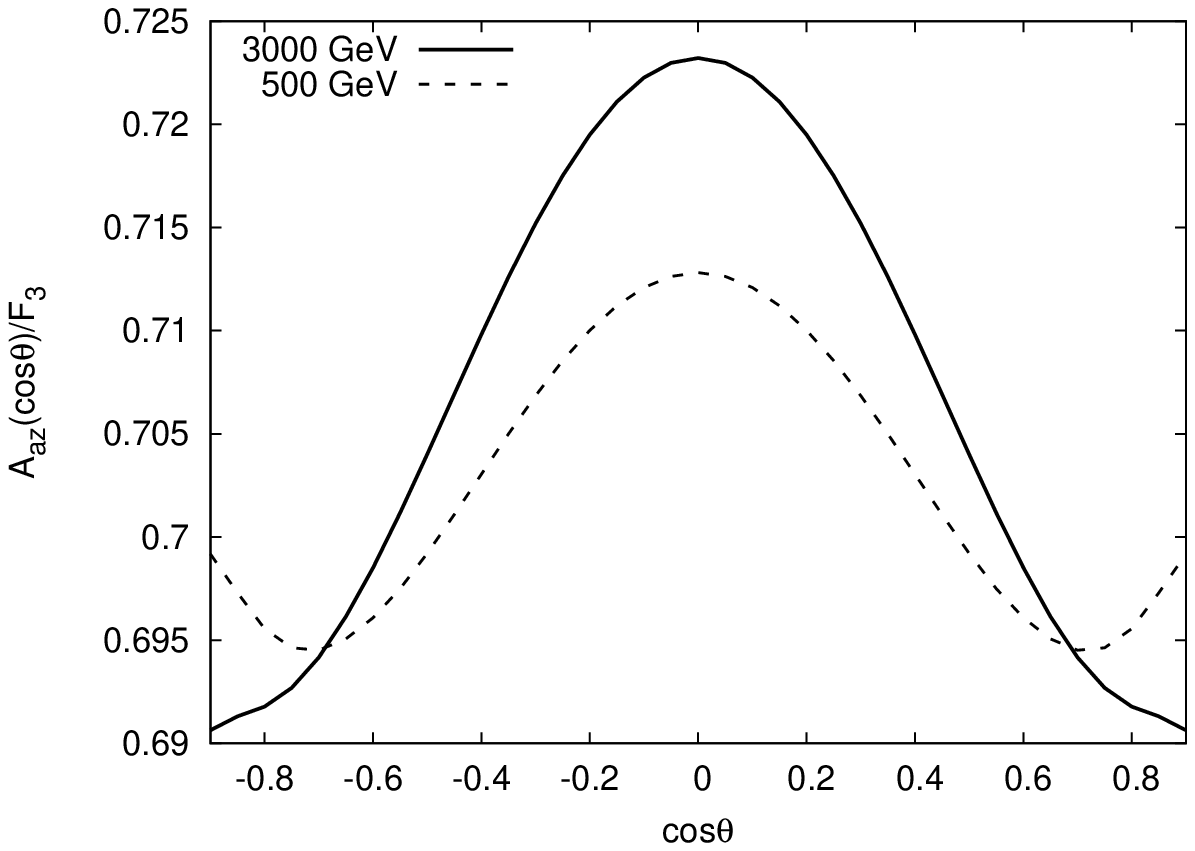}
\caption{Azimuthal asymmetry  $A_{az}(\cos\theta)/F_{3}$ as a function of polar angle
$\cos{\theta}$ between tau leptons and incoming photon direction in the center
of mass system of tau leptons.}
\label{fig3}}
   
Bounds on CP violating  anomalous electric  dipole  moment $d_{\tau}$   
can be found by $\chi^{2}$ analysis using azimuthal asymmetry  $A_{az}$
\begin{eqnarray}
\chi^{2}=&&\frac{(A_{az}(F_{3})-A_{az}(F_{3}=0))^{2}}
{\delta_{SM}^{2}}\\
\delta_{SM}=&&\sqrt{(\delta^{st})^{2}+(\delta^{sys})^{2}} \\
\delta^{st}=&&\frac{1}{\sqrt{N_{SM}}}\\
N_{SM}=&&L_{int} \sigma_{SM} BR 
\end{eqnarray}
where $\sigma_{SM}$, BR, $N_{SM}$ and $\delta_{SM}$ are  cross section, 
branching ratio, number of events 
and uncertainty without anomalous couplings.
$L_{int}$ is the integrated luminosity of CLIC.  $\delta^{st}$
and $\delta^{sys}$ are statistical and systematic uncertainties.
Table \ref{tab2}  shows the constraints
on the CP odd anomalous electric dipole  moment of the tau lepton 
which were computed using different CLIC luminosities 
and systematic uncertainties. Maximum virtuality of Weizsacker-Williams
photons  $Q_{max}^2=2$ $\mbox{GeV}^2$ has been taken. 

\TABLE{
\begin{tabular}{|c|c|c|c|c|c|}\hline
$L_{int}(fb^{-1})$ & $\sqrt{s}$ (GeV) & $|d_{\tau}|$(e cm) 
&$|d_{\tau}|$(e cm), $\delta_{1}^{sys}$ & $|d_{\tau}|$(e cm), 
$\delta_{2}^{sys}$ \\
\hline
50  & 500  & 3.34$\times 10^{-17}$ & 3.69$\times 10^{-17}$ & 1.59$\times 10^{-16}$   \\
100 & 500  & 2.36$\times 10^{-17}$ & 2.82$\times 10^{-17}$ & 1.57$\times 10^{-16}$    \\
230 & 500  & 1.56$\times 10^{-17}$ & 2.20$\times 10^{-17}$ & 1.56$\times 10^{-16}$    \\
\hline
100  & 1500  & 1.86$\times 10^{-17}$ & 2.50$\times 10^{-17}$ & 1.56$\times 10^{-16}$   \\
200 & 1500  & 1.32$\times 10^{-17}$ & 2.08$\times 10^{-17}$  & 1.56$\times 10^{-16}$    \\
320 & 1500  & 1.04$\times 10^{-17}$ & 1.90$\times 10^{-17}$ & 1.56$\times 10^{-16}$   \\
\hline
200  & 3000  & 1.16$\times 10^{-17}$ & 1.94$\times 10^{-17}$ & 1.56$\times 10^{-16}$  \\
400 & 3000  & 8.23$\times 10^{-18}$ & 1.76$\times 10^{-17}$  & 1.55$\times 10^{-16}$   \\
590 & 3000  & 6.77$\times 10^{-18}$ & 1.69$\times 10^{-17}$  & 1.55$\times 10^{-16}$   \\
\hline
\end{tabular}
\caption{Sensitivity of the process
$ee\to ee \tau^{+}\tau^{-}\to ee \pi^{+}\pi^{-}\nu\bar{\nu}$  to 
CP odd tau anomalous  electric
dipole moment $d_{\tau}$  at 95\% C.L. for CLIC energies
$\sqrt{s}=500, 1500, 3000$ GeV and integrated luminosities
given above. Total systematic uncertainties used in
$\chi^{2}$ function in the last two columns 
 have been taken as $\delta_{1}^{sys}=0.001$ and 
$\delta_{2}^{sys}=0.01$
\label{tab2}}}

The largest statistics of $\tau$ leptons have been collected by $e^{+}e^{-}$ 
B-factories (BELLE, BABAR) due to high cross section and luminosity. 
One of their strategy is to provide precision studies 
for basic $\tau$ properties as complementary to lower mass leptons. 
As given in the introduction, the most precise bounds on 
real and imaginary parts of  CP odd electric dipole form factor
were determined separately by BELLE collaboration \cite{belle} in the process
$e^{+}e^{-} \to \gamma \to \tau^{+}\tau^{-}$ with photon virtuality 
 $Q^2=100$ $GeV^2$. In our results average value of  virtuality $Q^2$ is pretty
smaller than 2 $GeV^2$.  
This means photons coupled to tau leptons are much 
close to real photons. Therefore CLIC  bounds of this work do 
not have one to one correspondence to BELLE bounds. For $\sqrt{s}=3000$ GeV
and systematic uncertainty less than $O(10^{-3})$, the limits 
obtained in Table \ref{tab2} can be considered to be 
complementary or  competitive  to the ones of BELLE. For systematic 
uncertainty larger than $O(10^{-2})$, statistics is dominated by systematic uncertainty. 
However, a firm decision can only be made after a detailed investigation 
of experimental conditions at CLIC. A theoretical study to find sensitivity of 
CLIC to electric dipole moment of $\tau$ lepton  has been done  recently 
in the paper \cite{alper} which contains comparable limits but only for CP even terms. 

\section{Discussion}
The main physics background for two photon collision case 
comes from the s-channel direct production of tau leptons.
Comparison of total cross sections has been made in the previous section 
for both cases, two photon induced t-channel production and s-channel direct production. 
Better way to discriminate them is to examine their transverse 
momentum distributions which is described briefly in Appendix B and is drawn 
in  Fig.\ref{fig4}. In this figure,  different
$p_{T}$ properties of two photon induced tau pair production cross
section $e^{+}e^{-} \to e^{+}e^{-} \tau^{-}\tau^{+}$  and
direct s-channel production $e^{+}e^{-} \to \tau^{-}\tau^{+}$ are given.
Remarkably dominant behaviour belongs to two photon collision for $p_{T} < 150$ GeV .
The upper curve increases up to a peak region around the half of the tau mass  
as $p_{T}$ approaches  origin and then  decreases down to zero when $p_{T}\to 0$.
This is connected to t-channel propagator where its denominator gets smaller as 
$p_{T}$ goes to origin. This is called collinear enhancement. When  
a factor of $p_{T}$ is combined with the propagator as in Eq.(B.1) (Appendix), 
the curve gives a maximum region about $p_{T}\simeq m/2$. If lepton mass vanishes
the collinear singularity occurs. In our case, tau mass takes the enhancement under 
control. In addition,  minimum transverse momentum cut of a few GeV (5 or more)
is used for efficient tau identification. This $p_{T}$ 
cut already excludes the peak region and the calculated 
sensitivities are safe.        
Another reason for overall increase of upper curve is based on the sum over 
$\sqrt{\hat{s}}$ photon spectrum where largest contributions of the 
subprocess $\gamma\gamma \to \tau^{+}\tau^{-}$  take place at 
small $\sqrt{\hat{s}}$ values. In the case of direct production only 
one energy value $\sqrt{s}=3000$ GeV contributes.   

\FIGURE{\epsfig{file=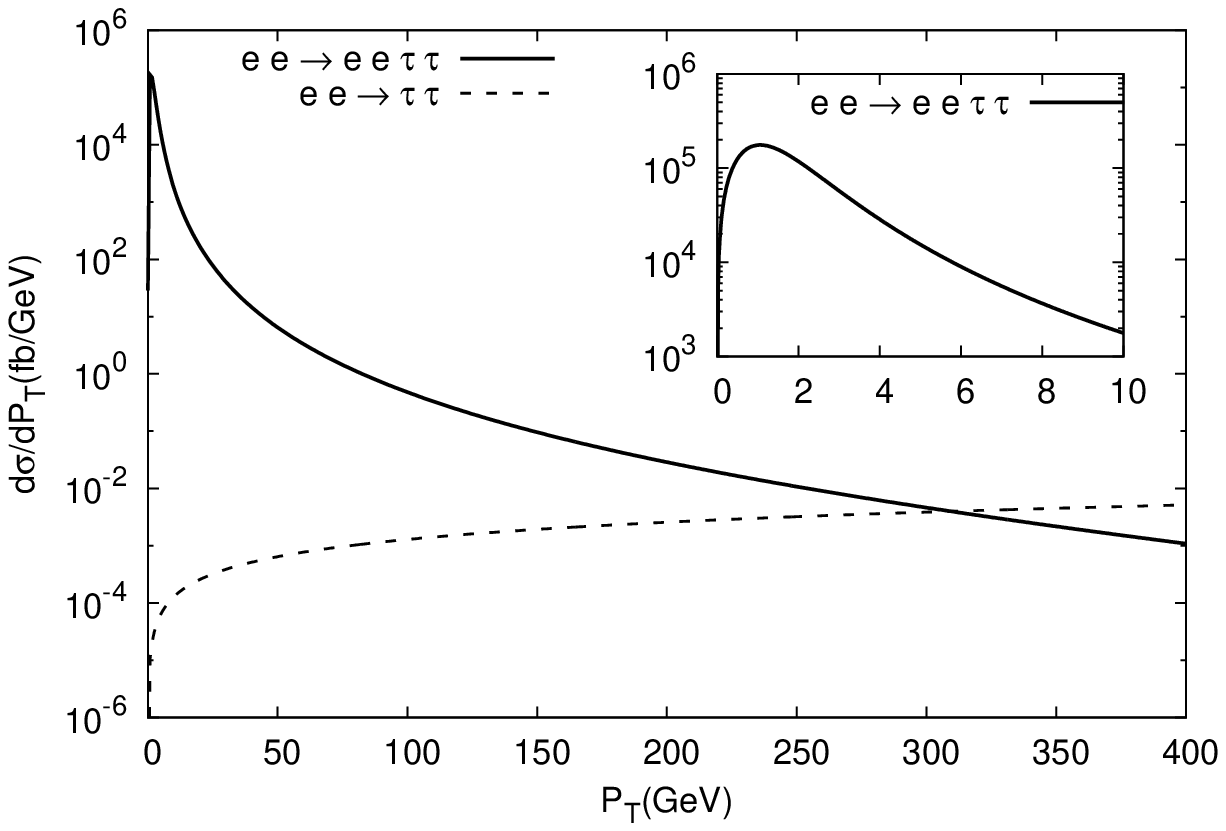}
\caption{Transverse momentum distributions of tau pair.
Two photon induced production of SM process with integration
over photon spectrum  (upper curve)
and direct tau pair production (lower curve)
at $\sqrt{s}=3000$ GeV. 
The same curve due to two photon induced production is shown inside the small frame
around tau mass region in order to see the typical behaviour that distribution goes 
to zero when $p_{T}\to 0$. }
\label{fig4}}

Another quantity is the invariant mass of tau pair  which differs
appreciably   in two cases.
In two photon induced process, invariant mass of  tau pair should be equal to
effective invariant mass of two photons  which is  less than 15-90 GeV. Invariant mass 
of tau pair in the s channel  case  is the same as incoming $e^{+}e^{-}$ center of 
mass energy  500-3000 GeV  with much smaller cross sections. 
As explained above, in small  $p_{T}$ region, background from direct production case 
can easily be eliminated.
In addition, if scattered electrons with small angles are tagged, 
two photon induced production 
cross section will have relatively clear background. 

An important point is to detect final tau leptons with an uncertainty 
as small as possible. This depends on the decay products of tau lepton  
beside the experimental tools.
In the subprocess $\gamma\gamma \to \tau^{-}\tau^{+}$, produced tau pair will 
not be back to back because of initial photon energies which are not 
the same with each other.  In hadronic tau decay $\tau\to \pi\nu$, 
charged pions will be detected with its momentum direction while 
neutrino escapes detection. For tau lepton energies  $m<< E$, the decay 
products will be collimated with opening angle of the order of  $m/E$.
It is interesting that opening angle is completely determined by the mass 
and energy of the hadron. Both of them are measurable and angle is calculated 
from the decay kinematics \cite{stahl}.  
When  scattered electrons    are tagged, initial photon energies can be 
determined by the difference of incoming  and outgoing beam energies 
$E_{\gamma}=E_{b}-E^{\prime}_{b}$. Then, the invariant mass of $\tau$ pair 
is obtained through photons momenta. Next, the $\tau$ decay products are boosted 
into the $\tau$ pair rest frame with boost parameter 
$(E^{\gamma}_{1}-E^{\gamma}_{2})/(E^{\gamma}_{1}+E^{\gamma}_{2})$
 where the energy of each $\tau$ is defined in terms of the 
invariant mass of tau pair. Therefore, the tau momentum can be constrained 
to be in a cone defined by  a fixed angle to the momentum of its hadronic decay product. 
Possible solutions for $\tau$ momenta correspond to intersections of two cones belongs 
to each $\tau$. In the case of events containing single neutrino from each $\tau$ decay, 
almost the same methods have been used for lepton colliders \cite{heister,belous}
 with  high efficiency without the need of knowledge 
about $\tau$ production vertex.

If deflected electrons and positrons are not required to be observed,
 the information about invariant 
mass of $\tau$ pair  is lost due to missing momentum of the neutrino. 
In this case, tau reconstruction is still possible 
if production vertex of $\tau$ pair  and  
charged prong trajectory are known sufficiently well. Let us consider the average 
 decay length of tau lepton   $L=\beta\gamma\tau_{\tau}$
in terms of  relativistic factor $\gamma=1/(\sqrt{1-\beta^{2}})$ and tau life time 
$\tau_{\tau}$. For the $\tau$ energy $E>20$ GeV 
the decay length reaches  mm size which leads to
 the measurable quantity using impact parameter method. Impact parameter 
is the shortest distance to the charged hadron momentum direction from the 
tau pair production vertex (or primary vertex). Impact parameter is crucial to 
determine secondary vertex ($\tau$ decay point) and  can be written 
approximately $d=L\psi$ in terms of small opening angle 
$\psi$ for sufficient $\tau$ energy. This 
method works well to determine neutrino momentum when single neutrino and 
single  charged pion are produced \cite{jeans}.   

In both cases above, full reconstruction of produced tau leptons is possible 
if single neutrino appears for each tau decay .   
To conclude, tau pair production in two photon collision at CLIC 
is sensitive to test CP odd EDM coupling  of tau lepton especially 
for $\sqrt{s}=3000$ GeV.

\appendix

\section{Definitions of Reduced Amplitudes}

Definitions of reduced amplitudes in Eqs. 2.3-2.5 :

\begin{eqnarray}
C_{0}=&&6m^{4}-2m^{2}(2\hat{s}+5\hat{t}+3\hat{u})+2\hat{t}\hat{u} \nonumber\\
 &&+F_{3}^2[\frac{1}{4m^2}((m^2-\hat{t})(9m^4+m^2(8\hat{s}-22\hat{t}
 +4\hat{u})+\hat{t}(9\hat{t}-4\hat{s})))]\\
D_{0}=&&6m^{4}-2m^{2}(2\hat{s}+3\hat{t}+5\hat{u})+2\hat{t}\hat{u}\nonumber\\
 &&+F_{3}^2[\frac{1}{4m^2}((m^2-\hat{u})(9m^4+m^2(8\hat{s}-22\hat{u}
 +4\hat{t})+\hat{u}(9\hat{u}-4\hat{s})))]\\
G_{0}=&&4(\hat{s}m^2-4m^4)\nonumber\\ &&+F_{3}^2[\frac{1}{8m^2}(-44m^6+m^4
 (32\hat{s}+34(\hat{t}+\hat{u}))\nonumber\\&&-2m^2(16\hat{s}^2 
 +8\hat{s}(\hat{t}+\hat{u})-9\hat{t}^2+20\hat{t}\hat{u}
  -9\hat{u}^2)+5\hat{s}^3\nonumber\\ &&+8\hat{s}^2(\hat{t}+\hat{u})
  -5\hat{s}(\hat{t}-\hat{u})^2-2(4\hat{t}^3+\hat{t}^2\hat{u}
  +\hat{t}\hat{u}^2+4\hat{u}^3))]\\
C_{xx}=&&4m^2(2k_{1x}^{2}+m^2+\hat{t}) \nonumber\\
 &&+F_{3}^2[\frac{1}{4m^2}(m^2(14k_{1x}^2\hat{s}-3\hat{s}^2+4\hat{s}\hat{t}
+3\hat{u}(\hat{u}-2\hat{t}))\nonumber \\ &&+\hat{t}(2k_{1x}^{2}\hat{s}
-\hat{s}^2+\hat{u}^2)-m^6+m^4(12\hat{s}+5\hat{t}-2\hat{u}))]\\
C_{yy}=&&4m^2(m^2+\hat{t})
 +F_{3}^2[\frac{1}{4m^2}(m^2(-3\hat{s}^2+4\hat{s}\hat{t}
+3\hat{u}(\hat{u}-2\hat{t}))\nonumber \\ &&+\hat{t}
(-\hat{s}^2+\hat{u}^2)-m^6+m^4(12\hat{s}+5\hat{t}-2\hat{u}))]\\
C_{zz}=&&4E^2(2k_{1z}^{2}-4k_{1z}p_{1z}+m^2+2p_{1z}^2+\hat{t})
 +4p_{1z}^2(m^2+\hat{t})\nonumber \\
 &&+F_{3}^2[\frac{1}{4m^4}\{E^2 (m^2(14k_{1z}^2\hat{s}-4k_{1z}p_{1z}
 (5\hat{s}+6\hat{t}-2\hat{u})\nonumber\\&&+p_{1z}^2(54\hat{s}
   -56\hat{t}+40\hat{u})
-3\hat{s}^2+4\hat{s}\hat{t}-6\hat{t}\hat{u}+3\hat{u}^2)\nonumber\\
 && +\hat{t}(2k_{1z}^2\hat{s}+4k_{1z}p_{1z}(\hat{s}+2\hat{u})
 +2p_{1z}^2(5\hat{s}+4\hat{u})\nonumber\\&&-\hat{s}^2+\hat{u}^2)
  +m^4(-56k_{1z}p_{1z}-56p_{1z}^2+12\hat{s}\nonumber\\
   &&+5\hat{t}-2\hat{u})-m^6)
 -p_{1z}^2(m^6+m^4(-12\hat{s}-5\hat{t}+2\hat{u})\nonumber\\&&+m^2(3\hat{s}^2
  -4\hat{s}\hat{t}+6\hat{t}\hat{u}-3\hat{u}^2)+\hat{t}
    (\hat{s}^2-\hat{u}^2))\}]\\
  C_{xy}^{-}=&&F_{3}[-4E(k_{1z}-p_{1z})(3m^2+\hat{t})]\\
  C_{yz}^{-}=&& F_{3}[-\frac{4E^2}{m}k_{1x}(3m^2+\hat{t})]\\
  C_{xz}^{+}=&& 8Ek_{1x}m(k_{1z}-p_{1z})\nonumber\\
  &&+F_{3}^2[\frac{1}{2m^3}\{Ek_{1x}(m^2(7k_{1z}\hat{s}-p_{1z}
  (\hat{s}+2\hat{t}-6\hat{u}))\nonumber\\&&+\hat{t}(k_{1z}\hat{s}+p_{1z}
  (\hat{s}+2\hat{u}))-22m^4p_{1z}\}]\\
D_{xx}=&&4m^2(2k_{1x}^{2}+m^2+\hat{u}) \nonumber\\
 &&+F_{3}^2[\frac{1}{4m^2}(m^2(14k_{1x}^2\hat{s}-3\hat{s}^2+4\hat{s}\hat{u}
+3\hat{t}(\hat{t}-2\hat{u}))\nonumber \\ &&+\hat{u}(2k_{1x}^{2}\hat{s}
-\hat{s}^2+\hat{t}^2)-m^6+m^4(12\hat{s}+5\hat{u}-2\hat{t}))]\\
D_{yy}=&&4m^2(m^2+\hat{u})
 +F_{3}^2[\frac{1}{4m^2}(m^2(-3\hat{s}^2+4\hat{s}\hat{u}
+3\hat{t}(\hat{t}-2\hat{u}))\nonumber \\ &&+\hat{u}
(-\hat{s}^2+\hat{t}^2)-m^6+m^4(12\hat{s}+5\hat{u}-2\hat{t}))]\\
D_{zz}=&&4E^2(2k_{1z}^{2}+4k_{1z}p_{1z}+m^2+2p_{1z}^2+\hat{u})
 +4p_{1z}^2(m^2+\hat{u})\nonumber \\
&&+F_{3}^2[\frac{1}{4m^4}\{E^2 (m^2(14k_{1z}^2\hat{s}+4k_{1z}p_{1z}
 (5\hat{s}+6\hat{u}-2\hat{t})\nonumber\\&&+p_{1z}^2(54\hat{s}
   -56\hat{u}+40\hat{t})
-3\hat{s}^2+4\hat{s}\hat{u}-6\hat{t}\hat{u}+3\hat{t}^2)\nonumber\\
 && +\hat{u}(2k_{1z}^2\hat{s}-4k_{1z}p_{1z}(\hat{s}+2\hat{t})
 +2p_{1z}^2(5\hat{s}+4\hat{t})\nonumber\\&&-\hat{s}^2+\hat{t}^2)
  +m^4(56k_{1z}p_{1z}-56p_{1z}^2+12\hat{s}\nonumber\\
   &&+5\hat{u}-2\hat{t})-m^6)
 -p_{1z}^2(m^6+m^4(-12\hat{s}-5\hat{u}+2\hat{t})\nonumber\\&&+m^2(3\hat{s}^2
  -4\hat{s}\hat{u}+6\hat{t}\hat{u}-3\hat{t}^2)+\hat{u}
    (\hat{s}^2-\hat{t}^2))\}]\\
  D_{xy}^{-}=&&F_{3}[4E(k_{1z}+p_{1z})(3m^2+\hat{u})]\\
  D_{yz}^{-}=&& F_{3}[-\frac{4E^2}{m}k_{1x}(3m^2+\hat{u})]\\
  D_{xz}^{+}=&& 8Ek_{1x}m(k_{1z}+p_{1z})\nonumber\\
  &&+F_{3}^2[\frac{1}{2m^3}\{Ek_{1x}(m^2(7k_{1z}\hat{s}+p_{1z}
  (\hat{s}+2\hat{u}-6\hat{t}))\nonumber\\&&+\hat{u}(k_{1z}\hat{s}-p_{1z}
  (\hat{s}+2\hat{t}))+22m^4 p_{1z}\}]\\
 G_{xx}=&&2(k_{1x}^2(8m^2-4\hat{s})+2m^4+4m^2(\hat{t}+\hat{u})
  +\hat{s}^2-\hat{t}^2-\hat{u}^2)\nonumber\\
 &&+F_{3}^2[\frac{1}{8m^2}(4k_{1x}^2(36m^4-10m^2(\hat{s}
 +4m^2)+3\hat{s}^2+7\hat{s}\hat{t}+7\hat{s}\hat{u}\nonumber\\
 &&+(\hat{t}+\hat{u})^2)
 -20m^6+10m^4(8\hat{s}+3\hat{t}+3\hat{u})+2m^2(4\hat{s}^2
 -8\hat{s}\hat{t}-8\hat{s}\hat{u}\nonumber\\
 &&-5\hat{t}^2-20\hat{t}\hat{u}
 -5\hat{u}^2)-3\hat{s}^3-4\hat{s}^2\hat{t}-4\hat{s}^2\hat{u}\nonumber\\
 &&+3\hat{s}\hat{t}^2+10\hat{s}\hat{t}\hat{u}+3\hat{s}\hat{u}^2
 +4\hat{t}^3+6\hat{t}^2\hat{u}+6\hat{t}\hat{u}^2+4\hat{u}^3)] \\
 G_{yy}=&&2(2m^4+4m^2(\hat{t}+\hat{u})+\hat{s}^2-\hat{t}^2-\hat{u}^2)
 \nonumber\\
 &&+F_{3}^2[\frac{1}{8m^2}(-20m^6+10m^4(8\hat{s}+3\hat{t}+3\hat{u})
 +2m^2(4\hat{s}^2-8\hat{s}\hat{t}-\hat{s}\hat{u}\nonumber\\
 &&-5(\hat{t}^2
 +4\hat{t}\hat{u}+\hat{u}^2)-3\hat{s}^3-4\hat{s}^2(\hat{t}-
 4\hat{s}^2\hat{u}+\hat{s}(3\hat{t}^2+10\hat{t}\hat{u}
 +3\hat{u}^2)\nonumber\\
 &&+4\hat{t}^3+6\hat{t}^2\hat{u}+6\hat{t}\hat{u}^2
 +4\hat{u}^3 )] \\
G_{zz}=&&\frac{2}{m^2}\{E^2(2k_{1z}^{2}(8m^2-4\hat{s})+8k_{1z}p_{1z}
 (\hat{t}-\hat{u})+2m^4\nonumber\\ &&+4m^2(6p_{1z}^2+\hat{t}+\hat{u})
 -12p_{1z}^2\hat{s}-8p_{1z}^2\hat{t}-8p_{1z}^2\hat{u}+\hat{s}^2
 -\hat{t}^2-\hat{u}^2)\nonumber\\ &&+p_{1z}^2(2m^4+4m^2(\hat{t}
  +\hat{u})+\hat{s}^2-\hat{t}^2-\hat{u}^2)\}\nonumber \\
&&+F_{3}^2[\frac{1}{8m^4}\{E^2 (4k_{1z}^2(36m^4-10m^2(3\hat{s}
  +2\hat{t}+2\hat{u})+3\hat{s}^2+7\hat{s}\hat{t}\nonumber\\
  &&+7\hat{s}\hat{u}
  +(\hat{t}+\hat{u})^2)-8k_{z}p_{z}(\hat{t}-\hat{u})(-8m^2 +\hat{s}
  +4\hat{t}+4\hat{u})-20m^6 \nonumber\\
  &&+m^4(80\hat{s}+30\hat{t}+30\hat{u}-304p_{1z}^2)
  -2m^2(p_{1z}^2(76\hat{s}+8\hat{t}+8\hat{u})\nonumber\\
  &&-4\hat{s}^2+8\hat{s}\hat{t}
  +8\hat{s}\hat{u}+5\hat{t}^2+20\hat{t}\hat{u}+5\hat{u}^2)\nonumber\\
  &&+36p_{1z}^2
  \hat{s}^2+76p_{1z}^2\hat{s}\hat{t}+76p_{1z}^2\hat{s}\hat{u}
  +28p_{1z}^2\hat{t}^2+24p_{1z}^2\hat{t}\hat{u}+28p_{1z}^2\hat{u}^2
  -3\hat{s}^3 \nonumber\\
  &&-4\hat{s}^2\hat{t}-4\hat{s}^2\hat{u}+3\hat{s}\hat{t}^2
  +10\hat{s}\hat{t}\hat{u}+3\hat{s}\hat{u}^2+4\hat{t}^3\nonumber\\
  &&+6\hat{t}^2\hat{u}+6\hat{t}\hat{u}^2+4\hat{u}^3)
  +p_{1z}^2(-20m^6+10m^4(8\hat{s}+3\hat{t}+3\hat{u})\nonumber\\
  &&+2m^2(4\hat{s}^2-8\hat{s}\hat{t}-8\hat{s}\hat{u}-5\hat{t}^2
  -20\hat{t}\hat{u}-5\hat{u}^2)-3\hat{s}^3-4\hat{s}^2\hat{t}
  -4\hat{s}^2\hat{u}
   \nonumber\\&&+\hat{s}(3\hat{t}^2+10\hat{t}\hat{u}
  +3\hat{u}^2)+4\hat{t}^3+6\hat{t}^2\hat{u}+6\hat{t}\hat{u}^2
  +4\hat{u}^3)\}] \\
  G_{xy}^{-}=&&F_{3}[4E(k_{1z}(\hat{t}-\hat{u})+4m^2p_{1z}
  +p_{1z}(5\hat{s}+2\hat{t}+2\hat{u}))]\\
  G_{yz}^{-}=&& F_{3}[\frac{4E^2}{m}k_{1x}(\hat{t}-\hat{u})]\\
  G_{xz}^{+}=&& \frac{8E}{m}(k_{1x}(2k_{1z}m^2-k_{1z}\hat{s}+p_{1z}\hat{t}
  -p_{1z}\hat{u})\nonumber\\
  &&+F_{3}^2[\frac{1}{2m^3}\{Ek_{1x}(k_{1z}(36m^4-10m^2(3\hat{s}+2\hat{t}
  +2\hat{u})+3\hat{s}^2+7\hat{s}(\hat{t}+\hat{u})\nonumber\\
  &&+(\hat{t}+\hat{u})^2 )+3p_{1z}\hat{s}(\hat{t}-\hat{u}))\} 
\end{eqnarray}
where  $k_{1}$, $k_{2}$, $p_{1}$ and $p_{2}$  are the 
momenta of the incoming photons and final $\tau$ leptons. 
Mandelstam variables are defined as $\hat{s}=(k_{1}+k_{2})^{2}$,
$\hat{t}=(k_{1}-p_{1})^{2}$ and $\hat{u}=(k_{1}-p_{2})^{2}$. 
E and m are energy and mass of tau leptons. 

\section{Transverse Momentum Distribution}

Transverse momentum $p_{T}$ distribution  of final tau leptons in the center of mass 
system of tau pair  is calculated using the following expressions :
\begin{eqnarray}
\frac{d\hat{\sigma}}{dp_{T}}&&=\frac{p_{T} |M|^{2} \delta(y_{\tau}-y_{0})}{8\hat{s}\pi
(2\sqrt{\hat{s}} E_{T}\sinh{y_{0}})}dy_{\tau} \\
y_{0}&&=\mbox{arccosh}{(\frac{\sqrt{\hat{s}}}{2E_{T}})} \\
E_{T}&&=\sqrt{p_{T}^{2}+m^{2}} 
\end{eqnarray}
where $y_{\tau}$ is the rapidity of tau lepton:
\begin{eqnarray}
y_{\tau}=\frac{1}{2} \ln{\frac{E+p_{z}}{E-p_{z}}}.
\end{eqnarray}
Mandelstam variables in the Spin averaged square of the Feynman amplitude $|M|^2$ 
can also be   written  in terms of $p_{T}$ and  $y_{\tau}$
\begin{eqnarray}
\hat{t}&&=m^2-\sqrt{\hat{s}} E_{T} e^{-y_{\tau}} \\
\hat{u}&&=m^2-\sqrt{\hat{s}} E_{T} e^{y_{\tau}}
\end{eqnarray}
 For the case of two photon collision two more integrations over photon 
spectrum are needed.

\end{document}